\documentclass[structabstract]{aa}
\usepackage{txfonts}
\usepackage{natbib}
\bibpunct{(}{)}{;}{a}{}{,} 
\usepackage{graphicx}
\usepackage{url}

\newcommand{\teff}{\mbox{$T_{\rm eff}$}}
\newcommand{\logg}{\mbox{$\log{g}$}}
\newcommand{\feh}{\mbox{$[Fe/H]$}}
\newcommand{\afe}{\mbox{$[\alpha/Fe]$}}

\newcommand{\Msun}{\mbox{$M_{\odot}$}}

\newcommand{\phx}{PHOENIX}

\begin{document}

\title{A new extensive library of PHOENIX stellar atmospheres and synthetic spectra}

\author{Tim-Oliver Husser\inst{1}
  \and Sebastian Wende - von Berg\inst{1}
  \and Stefan Dreizler\inst{1}
  \and Derek Homeier\inst{1}\inst{2}
  \and Ansgar Reiners\inst{1}
  \and Travis Barman\inst{3}
  \and Peter H. Hauschildt\inst{4}
}

\institute{Institut f\"ur Astrophysik, Georg-August-Universit\"at G\"ottingen, Friedrich-Hund-Platz 1, 37077 G\"ottingen, Germany
  \and CRAL, UMR 5574, CNRS, Universit\'e de Lyon,  \'Ecole Normale Sup\'erieure de Lyon, 46 All\'ee d'Italie, F-69364 Lyon Cedex 07, France
  \and Lowell Observatory, 1400 West Mars Hill Road, Flagstaff, AZ, 86001, USA
  \and Hamburger Sternwarte, Gojenbergsweg 112, 21029 Hamburg, Germany}

\date{Received / Accepted }

\abstract 
{} 
{We present a new library of high-resolution synthetic spectra based on the stellar atmosphere code \phx\ that can
be used for a wide range of applications of spectral analysis and stellar parameter synthesis.} 
{The spherical mode of \phx\ was used to create model atmospheres and to derive detailed synthetic stellar
spectra from them. We present a new self-consistent way of describing micro-turbulence for our model 
atmospheres.} 
{The synthetic spectra cover the wavelength range from $500\,\mbox{\AA}$ to $5.5\,\mathrm{\mu m}$ with resolutions of $R=500\,000$ in the
optical and near IR, $R=100\,000$ in the IR and $\Delta\lambda = 0.1\,\mbox{\AA}$ in the UV. The
parameter space covers $2\,300\,\mathrm{K} \leq \teff \leq 12\,000\,\mathrm{K}$, $0.0 \leq \logg \leq +6.0$,
$-4.0 \leq \feh \leq +1.0$, and $-0.2 \leq \afe \leq +1.2$. The library is a work in progress
and we expect to extend it up to $\teff=25\,000\,\mathrm{K}$.}
{}


\keywords{Convection - Radiative transfer - Stars: abundances - Stars: atmospheres}
\maketitle 

\section{Introduction}
Stellar spectral libraries are an important tool for analysing observed spectra and can 
be used for a wide range of applications. 
The use of empirical libraries guarantees real physical spectra, which will match
other observations well, e.g. absorption lines created by atomic and molecular transitions
appear at the exact position with the correct shapes and strengths. However,
these spectra are limited both in spectral resolution and wavelength coverage -- and extensions 
of the library require some
significant effort and observing time, if possible at all (e.g. changes
in instrumentation). Furthermore, the chemical composition
of the observed stars is not known exactly, resulting in possible systematic errors for
abundance determinations, and the spectra cover only a limited parameter space, 
usually restricted to the effective temperature, the
surface gravity, and the metallicity. Examples for
empirical libraries covering the optical wavelength range are the
Indo-U.S. Library of Coud\'e Feed Stellar Spectra
\citep{2004ApJS..152..251V}, MILES \citep{2006MNRAS.371..703S}, STELIB
\citep{2003A&A...402..433L}, and ELODIE
\citep{2004astro.ph..9214P}. Note that all these libraries contain
only low to medium resolution spectra.

Synthetic spectra on the other hand are limited by the completeness of the spectral line lists and 
by the knowledge of line broadening parameters and numerical assumptions -- e.g. plane-parallel 
versus spherical geometry and local thermodynamic 
equilibrium (LTE) versus statistical equilibrium (NLTE). Nevertheless, synthetic libraries
have many advantages: the range of stellar parameters, elemental abundances, and both the wavelength range
and the spectral resolution can be adjusted as needed. The best known set of model 
atmospheres currently available was created by \cite{1979ApJS...40....1K}, from which many
spectral libraries have been derived (e.g. \cite{2005ApJ...626..411R} in the UV, and \cite{2004MNRAS.351.1430M}
and \cite{2005A&A...442.1127M} in the full optical domain). The NLTE stellar atmosphere code TLUSTY 
\citep{1988CoPhC..52..103H} made possible a grid of NLTE atmospheres for O stars \citep{2003ApJS..146..417L}. \cite{2008A&A...486..951G} published
a grid of spectra based on MARCS which, like our new grid, uses spherical-symmetry for its models.

Because we need a stellar spectral library for the analysis of spectra observed with
MUSE \citep{2010SPIE.7735E...7B}, X-shooter \citep{2011A&A...536A.105V}, and CRIRES
\citep{2004SPIE.5492.1218K} for example, a decision was made in favour of the flexibility of a synthetic library that
can be extended on the fly -- even in parameters that are not changed in the published grid. For example, varying 
the abundances of single elements enables us to determine the areas of the spectra that are mainly affected by those elements. 
In comparison with empirical spectra, this shows a major advantage of having a stellar atmosphere code available 
together with thousands of synthetic atmospheres that can be used as starting points for extensions. Instead of extending an existing synthetic 
library, we opted for creating a new one using the latest state-of-the-art version of the stellar atmosphere code 
\phx\ \citep{1999JCoAM.109...41H}.

The first published \phx\ library consisted of 700 model atmospheres for M (sub)dwarf stars \citep{1995ApJ...445..433A},
all calculated under the assumption of local thermodynamic equilibrium. This has also been the case
for the NextGen library \citep{1999ApJ...512..377H} with its temperatures ranging from 3000\,K to 10\,000\,K, which 
has been used as the starting point for a grid of NLTE model atmospheres of dwarfs and giants.
Furthermore, an upgrade to spherically symmetric models for giant stars with effective temperatures between 3000\,K and 6800\,K
has been done by \cite{1999ApJ...525..871H}. Further calculations of NLTE models have been conducted by \cite{2000ApJ...537..946B}.
The effects of dust settling in cool stars like brown dwarfs have been investigated by \cite{2001ApJ...556..357A} and extended to
extra-solar planets by \cite{2009A&A...506.1367W,2011A&A...529A..44W}.
The latest comprehensive grid of spectra \citep{2005ESASP.576..565B} was created for use with the GAIA mission \citep{2008IAUS..248..217L}. 
In the following we will refer to spectra created with that version of \phx\ as AMES-cond-v2.6.

Our new spectral library is the first one using the current version 16 of \phx, which has a new equation of state as well as 
an up-to-date atomic and molecular line list. This allowed us to produce spectra that match observations 
especially of cool stars significantly better than other synthetic libraries. A full discussion of this 
aspect will appear in a subsequent publication \citep{WendeInPrep}.
The use of spherical geometry guarantees a consistent model grid from the main sequence up to giants. 
We have adjusted the wavelength range and resolution to meet the requirements for science cases using existing and upcoming
instruments.

In this paper we will first describe the basic parameters of the grid, namely its dimensions and the resolution
of the spectra. Then we will discuss in detail some of the input parameters such as the used element abundances, 
the stellar mass, and the micro-turbulence for which we will also provide comparisons with observed
data. We will characterise the equation of state used for calculating the models and show the improvements
that we achieved compared to older \phx models, other spectral libraries, and observations. 
Finally, we will explain how to download the FITS files containing the spectra and describe their content.

\section{The library}

\subsection{Dimensions and resolution}
Since the library is still a work in progress, we only present its current state which, however,
already covers the parameter space of most stellar populations that are not part of starburst regions. Table~\ref{table:paramspace} 
shows the current parameter space of the grid, a future extension will probably be towards
higher effective temperatures, most probably up to $25\,000\,\mathrm{K}$.
The grid is complete in its first three dimensions $\teff$, $\logg$ [cgs] and $\feh$. So far, a variation
of the alpha element abundance ($\afe \neq 0$) has only been calculated for effective temperatures
$3\,500\,\mathrm{K} \leq \teff \leq 8\,000\,\mathrm{K}$ and metallicities $-3 \leq \feh \leq 0$.

The standard spectroscopic abundance notation is used in this article which defines the ratio of two elements
$A$ and $B$ in a star relative to their ratio in the Sun as
\begin{equation}
  [A/B] \equiv \log [n(A)/n(B)] - log [n(A)/n(B)]_\odot,
\end{equation}
with the number density $n$.
We assume that $[X/H]=\feh$ for most elements and so, $\feh$ denotes the overall metallicity. 
The abundances of the $\alpha$ elements (O, Ne, Mg, Si, S, Ar, Ca, and Ti) are defined as
\begin{equation}
  [\alpha/H] = \feh + \afe.
\end{equation}
This implies that the value of \feh\ is conserved in alpha-enhanced or depleted models, but the overall 
metallicity Z changes.

\renewcommand{\tabcolsep}{0.5mm}
\begin{table}
  \caption{Parameter space of the grid. Alpha element abundances $\afe \neq 0$ are only available for $3\,500\,\mathrm{K} \leq \teff \leq 8\,000\,\mathrm{K}$ 
	   and $-3 \leq \feh \leq 0$.}
  \label{table:paramspace}
  \centering                                      %
  \begin{tabular}{lrrlc}
    \hline \hline
    Variable		   & & \multicolumn{2}{c}{Range}  & Step size \\ \hline
    $\teff$ [K] & & 2\,300 & -- 7\,000              & 100 \\
			   & & 7\,000 & -- 12\,000              & 200 \\
    $\logg$         	   & &   0.0 & -- +6.0     & 0.5 \\
    $\feh$               & &  -4.0 & -- -2.0    & 1.0 \\
		           & &  -2.0 & -- +1.0    & 0.5 \\
    $\afe$           & &  -0.2 & -- +1.2     & 0.2 \\ \hline
  \end{tabular}
\end{table}
\renewcommand{\tabcolsep}{2mm}



\renewcommand{\tabcolsep}{0.5mm}
\begin{table}
  \caption{Sampling of the spectra in the grid.}
  \label{table:resolution}
  \centering                                      %
  \begin{tabular}{rlrl}
    \hline \hline
    \multicolumn{2}{c}{Range [\AA]}      & \multicolumn{2}{c}{Sampling} \\ \hline
        500 & -- 3\,000    & $\Delta\lambda$ & = $0.1\mbox{\AA}$ \\
     3\,000 & -- 25\,000   & $R$ & $\approx 500\,000$ \\
    25\,000 & -- 55\,000   & $R$ & $\approx 100\,000$ \\ \hline
  \end{tabular}
\end{table}
\renewcommand{\tabcolsep}{2mm}
The spectra cover the whole wavelength range from $500\,\mbox{\AA}$ up to $5.5\,\mathrm{\mu m}$; the sampling is
given in Table~\ref{table:resolution}. Please note that these sampling rates are the direct output of \phx, i.e.
the spectra have never been resampled or convolved with any kernel.

The wavelength grid is identical for all models in the grid, which was made possible by a new option in PHOENIX v16, which
has been introduced in this new version and allows us to disable the creation of new wavelength points to better account for opacity.
We activated this option, since the sampling rate is high enough already. Furthermore, it allows us to store the
wavelength grid only once for all spectra in the library and therefore save storage space.

\subsection{Calculation of model atmospheres and synthetic spectra}
Synthesising a spectrum with \phx\ is a three-step process. First, a model atmosphere
was calculated using a set of 766 atmospheres from a previously unpublished \phx\ grid as start values. 
From those, the grid was extended step by step, always using an existing neighbouring model as the starting point. The
\phx\ code iterates a model atmosphere, until its criteria for convergence are reached. In order to
limit the run time, we stopped after every 80 iterations and restarted \phx\
if necessary. Parameter adjustments for many of the models had to be done individually in order
to obtain fully converged model atmospheres.

The next step was to activate a mode in \phx\ that triggers the use of special
line profiles, which we used for the Ca lines, for example. In this mode no convergence is checked,
so we carried out five more iterations by default in order to stratify the atmosphere again properly. Finally, 
the high resolution spectrum was synthesised from this intermediate model.

Having access to the model atmospheres as well as the spectra allows us to quickly expand
the grid in any given direction. Furthermore, we can use the existing model atmospheres for other
investigations -- e.g. we activated the output of the full radiative field, which can be used for calculating 
limb darkening coefficients in \phx\ for some models. The results will be discussed in a subsequent 
publication.

Even in its current state (end of 2012) with $\sim$50\,000 model atmospheres and corresponding synthetic spectra
(of which $\sim$30\,000 will be published), it took about $\sim$135 CPU years for the new \phx\ library to be 
calculated on the Nehalem Cluster of the GWDG\footnote{Gesellschaft f\"ur 
wissenschaftliche Datenverarbeitung mbH G\"ottingen, \url{http://www.gwdg.de/index.php?id=2156}.} in about one year,
parallelised on 20-50 nodes with 8 CPUs each.

\subsection{\phx\ settings}
While most other synthetic libraries set many parameters to fixed values, we tried to use parametrisations,
for the mass and the mixing length parameter $\alpha$, for example, that better match observations. Below we
introduce a new unique method for the micro-turbulence, which we calculate from convection velocities
obtained from the atmospheric model. Furthermore our new stellar library is the first one to use 
the new equation of state introduced with the new version of \phx\ and we use the most recent values for the
solar element abundances.

All model atmospheres have been calculated in a spherical mode with 64 layers. The reference wavelength 
defining the mean optical depth grid, which should be set to a wavelength with sufficient
flux, is fixed to $\lambda_\tau=12\,000\,\mbox{\AA}$ for $\teff<5\,000\,\mathrm{K}$ and $\lambda_\tau=5\,000\,\mbox{\AA}$ 
for hotter stars.

Local thermodynamic equilibrium (LTE) has been assumed for all the models in the library, which is sufficient
for the temperature range that we have investigated so far -- the extension to effective temperatures $\teff \ge 12\,000\,\mathrm{K}$ 
will take into account NLTE effects. Nevertheless, even for the existing models with
$\teff \geq 4\,000\,\mathrm{K}$, we used the NLTE mode of \phx\ to use special line profiles for some species 
(Li I, Na I, K I, Ca I, Ca II). The full sample of molecules is only used for models with $\teff \leq 8\,000\,\mathrm{K}$,
while it is stripped to a few important ones for hotter stars.

Condensation is included in the equation of state but ignored in the opacity \citep[Cond setup, see][]{2001ApJ...556..357A}
and we do not handle dust settling at all in our model atmospheres, since all our models have effective temperatures 
$\teff \geq 2\,300\,\mathrm{K}$.

\subsubsection{Mass}
While plane-parallel geometry is a valid assumption for main sequence star model atmospheres, this is no longer true for giants.
For those, \phx\ offers a mode for spherical symmetry, which we used for all models to guarantee 
a homogeneous grid. A model atmosphere with spherical geometry is then described by its effective temperature $\teff$, surface gravity $\logg$,
and mass $M_\star$, from which all the other physical parameters can be derived, i.e. adjusting the mass changes both the radius 
and the luminosity. In order to get numerically stable and realistic atmospheres, we used a varying mass throughout the 
grid and did not fix it to specific values (as \cite{1999ApJ...525..871H} and \cite{2008A&A...486..951G} do, for example).

For main sequence stars we use the mass-luminosity relation, which can be approximated for solar type stars \citep{Voigt} as
\begin{equation}
  L_\star/L_\odot = (M_\star/M_\odot)^3.
\end{equation}
Furthermore we can calculate the luminosity from the effective temperature and radius. For the Sun the radius can 
be substituted by the surface gravity and mass, which yields:
\begin{equation}
L_\star/L_\odot = (T_{\mathrm{eff},\star} /T_{\mathrm{eff},\odot})^4 (R_\star/R_\odot)^2 
		 = (T_{\mathrm{eff},\star} /T_{\mathrm{eff},\odot})^4 M_\star/M_\odot.
\end{equation}
By combining these two equations we get:
\begin{equation}
 (T_{\mathrm{eff},\star} /T_{\mathrm{eff},\odot})^4 = (M_\star/M_\odot)^2.
\end{equation}

For evolved stars, we can naturally not prescribe a unique mass-radius relation,
and it becomes more challenging to define mass as a function of \teff\ and \logg, 
so we modify the main sequence relation with a variable factor $c$ for lower surface
gravities to match observed values of giants and super giants.
Masses, especially of red giants and asymptotic giant branch stars, show considerable
spread even for identical atmospheric parameters, but we chose adjustment factors
such as matching the very-well-constrained values of Arcturus \citep{2011ApJ...743..135R}
and the sample of M giants studied by \citet{1998NewA....3..137D}, roughly covering 
$3000\,\mathrm{K} \le \teff \le 4000\,\mathrm{K}$, $-0.5 \le \logg \le 1.5$ with masses 
between 0.8 and 4\,\Msun. Therefore in the grid the mass is given by
\begin{equation}
  M_{\star} = c \cdot M_{\sun} \cdot \left( \teff / 5\,770\,\mathrm{K} \right)^2,
  \label{eq:mass}
\end{equation}
with values for the coefficient given in Table~\ref{table:massParams}.
\begin{table}
  \caption{Coefficients for calculating the mass in Eq.~\ref{eq:mass}.}
  \label{table:massParams}
  \centering      
  \begin{tabular}{r|ccccccc}
    \hline \hline
    log(g) & $>4$  & $>3$  & $>2$  & $>1.6$  & $>0.9$  & $>0$  & $\leq0$ \\
    c      & 1     & 1.2   & 1.4   & 2       & 3       & 4     & 5 \\ \hline
  \end{tabular}
\end{table}
\begin{figure}
  \resizebox{\hsize}{!}{\includegraphics{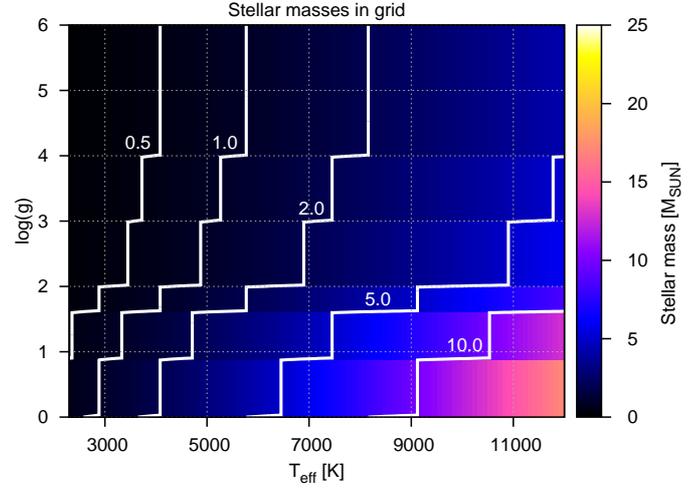}}
  \caption{Distribution of stellar masses for different effective temperatures $\teff$ and surface gravities $\logg$, 
           colour-coded in units of solar mass from $0\,M_\odot$ (black) to $25\,M_\odot$ (white).}
  \label{figure:mass}
\end{figure}
Figure~\ref{figure:mass} shows the distribution of masses over the grid.
For solar-like stars, we obtain $M_\star=1M_\odot$ and the masses increase for
hotter and larger stars as expected.

\subsubsection{Convection}
In \phx, convection in the stellar atmosphere is described by the mixing length theory
of turbulent transport, which goes back to \cite{Prandtl} and, in the stellar context, to 
\cite{1953ZA.....32..135V} and \cite{1958ZA.....46..108B}. It is characterised by the  
mixing length parameter $\alpha$, which describes the ratio between the
characteristic length that a volume of gas can rise in a stellar atmosphere before mixing with its surrounding
and the pressure scale height ($\alpha=l/H_p$), i.e. it is a value for the efficiency of the 
convective energy transport -- a large $\alpha$ indicates an efficient energy transport, while it is inefficient
for a small $\alpha$ in which case the gas volume is allowed to rise to greater heights under the buoyant force, 
thus reaching higher velocities and carrying more heat before it is assumed to be dissipated. 

For our spectra we describe the mixing length parameter $\alpha$ as discussed in \cite{1999A&A...346..111L}
where it has been calibrated using 3D radiative hydrodynamic models. Since the authors emphasise that it might be incorrect to extrapolate the mixing length
outside the ranges given in the paper ($4\,300\,\mathrm{K} \leq \teff \leq 7\,100\,\mathrm{K}$ and $2.54 \leq \logg \leq 4.74$),
we restrict the result to an interval $\alpha \in [1,3.5]$.

\begin{figure}
  \resizebox{\hsize}{!}{\includegraphics{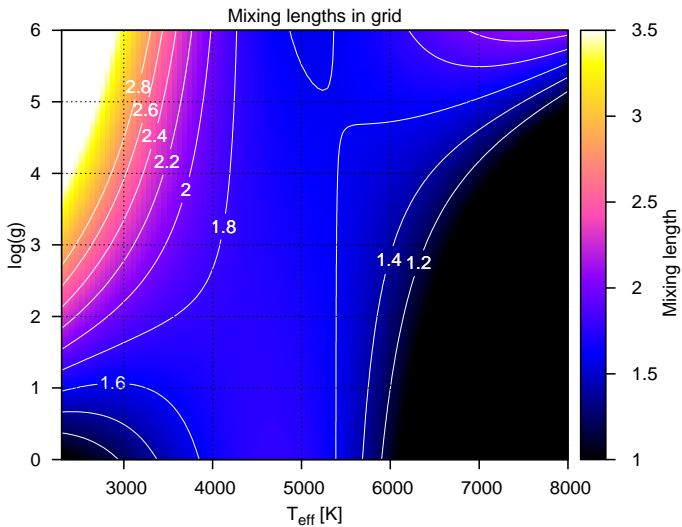}}
  \caption{Distribution of the mixing length parameter $\alpha$ in the grid.}
  \label{fig:mixingLength}
\end{figure}
The distribution of mixing lengths in the grid is shown in Fig.~\ref{fig:mixingLength}. The majority of the 
models have an $\alpha$ roughly between 1.5 and 2.5, while we had to cut off at $\alpha=3.5$ for models with 
$\teff \lessapprox 3\,000\,\mathrm{K}, \logg \gtrapprox 3$ and at $\alpha=1$ for those with 
$\teff \gtrapprox 6\,000\,\mathrm{K}, \logg \lessapprox 5$. For almost all models with 
$\teff \gtrapprox 8\,000\,\mathrm{K}$ the mixing length is $\alpha=1$.

Because all the other changes introduced with the new \phx\ library, we unfortunately are unable
to comment on the impact of our choice of this varying alpha parameter on the final spectra.

\subsubsection{Micro-turbulence}
\begin{figure*}
  \centering
  \includegraphics{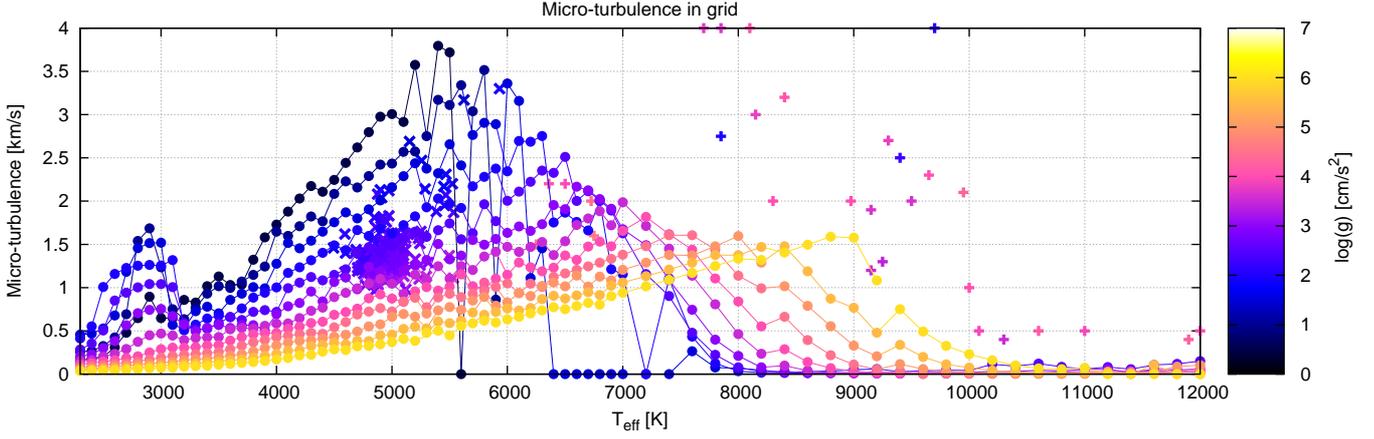}
  \caption{Distribution of micro-turbulence for models with solar abundances for different
           effective temperatures $\teff$ and surface gravities $\logg$ (colour-coded). We had to disable the convection
           for the giants with $\teff \gtrsim 5\,500$\,K  and therefore no micro-turbulence
           was applied to these spectra. In and around this area we observe numerical instabilities causing
           single models to show differing results. Over-plotted are observational data points taken from \cite{2009A&A...503..973L} (+) 
           and \cite{2008PASJ...60..781T} (x).}
  \label{figure:microTurb}
\end{figure*}

Micro-turbulence is an ad-hoc parameter that is used to match synthetic spectra with observations. We use it in the sense that the large-scale 
(macro-) turbulent motion triggers small-scale (micro-) turbulent motion below the length scale of the mean free path of the photons. 
Unlike macro-turbulent motion, this motion affects the strength of a spectral line \citep{gray2005observation}, 
since it directly influences the line opacities, so it has to be included 
in the line forming process and cannot be applied afterwards. Following this scheme, micro-turbulence is 
strongly related to the macro-turbulent motion and we therefore use the relation
\begin{equation}
  v_\mathrm{micro} = 0.5 \cdot \left< v_\mathrm{conv} \right>
\end{equation}
to determine its amount. This empirical approximation follows from 3D radiative hydrodynamic investigations of cool M stars 
\citep{2009A&A...508.1429W}, but also gives reasonable values for hotter stars (see Fig.~\ref{figure:microTurb}). The mean value of the macro-turbulence
$\left< v_\mathrm{conv} \right>$ is taken from the \phx\ atmospheres prior to spectral synthesis. The 
convective velocities from all layers are averaged in a way that only layers with non-zero values 
are included and then divided by a factor of two, i.e. micro-turbulence is assumed to be half of the mean convective 
velocity in the photosphere.

Since micro-turbulence mainly affects the strength and shape of the lines and not the structure of
the atmosphere, it is only included in the computation of the high resolution synthetic spectra and not in the underlying model 
atmospheres. We did not attempt to further converge the atmospheres with the micro-turbulences derived from 
the previous iterations, which would have cost additional computational resources without 
noticeably changing the structure. 
Despite the lack of a self-consistent description derived from first principles, this empirical description removes micro-turbulence as a free parameter.

Figure~\ref{figure:microTurb} shows the distribution of micro-turbulences for models with solar abundances. As one can see, for cool
stars our micro-turbulences are a lot smaller that the 2\,km/s 
frequently assumed in model atmospheres. 
Over-plotted are observational data points taken from \cite{2009A&A...503..973L} and 
\cite{2008PASJ...60..781T}. 
The agreement with our values for stars on the upper main-sequence and the 
red giant branch is gratifying. For effective  temperatures above 6\,500\,K, however,
convective turbulence predicted by our models decreases steadily and cannot 
reproduce the significant broadening velocities observed by \citet{2009A&A...503..973L}
in F and A stars. 

We note however that the \citet{2009A&A...503..973L} study does not display a unique 
relation of photospheric line broadening and atmospheric parameters, and includes a 
number of peculiar stars -- e.g. all main sequence stars with $v_\mathrm{micro} \ge 3\,\mathrm{km/s}$ 
in their sample are Am stars. They also conclude that 
the line shapes of these stars cannot be understood in a simple micro-turbulence concept. 
These findings are essentially confirmed by RHD simulations 
\citep[and references therein]{2012JCoPh.231..919F}, but these still sample the hottest stars 
with convectively unstable atmospheres only very sparsely. 

For giant stars with $\teff \approx 6\,500\,\mathrm{K}$
we encountered strong numerical instabilities in the \phx\ convection module as a consequence of 
to the diminishing extension of the convectively unstable layer and had to disable it completely. 
For consistency with the rest of our grid, we have chosen to publish these spectra with 
zero $v_\mathrm{micro}$; the users are advised that they cannot replace more detailed 
modelling using the micro-turbulence as a free parameter for those stars.

Other unexpected results appear in the region below 3\,000\,K, where we see another peak in the 
micro-turbulence. Those models develop a second convective zone at the outer edge of the 
atmosphere, where the convection velocity increases towards lower optical depths, which causes our 
method to produce values for the micro-turbulence that are higher than expected. A similar behaviour
has been observed in 3D radiative hydrodynamic simulations by \cite{2009A&A...508.1429W}.
Although the effect of this second convective zone on the spectrum is small, it does introduce
changes as a result of our new self-contained method for calculating the micro-turbulence. Thus,
we will provide an update for the spectral library for stars with $\teff \lessapprox 3\,000\,\mathrm{K}$
within a couple of months, in which we will limit the range of optical depths used for calculating
the micro-turbulence.

\subsubsection{Equation of state}
The Astrophysical Chemical Equilibrium Solver \citep[ACES,][]{Barman2012} equation of 
state (EOS) that is used starting with \phx\ version 16 is a state-of-the-art 
treatment of the chemical equilibrium in a stellar atmosphere. It uses
the method of \cite{smith1982chemical} with new experimental and theoretical 
thermodynamical data (Barman 2012) for 839 species (84 elements, 289 ions, 249 molecules,
217 condensates).

In each layer, the chemical equilibrium for all used atomic and molecular species is computed in dependence on the pressure, temperature, 
and density. After computing the atmospheric structure with the equations of radiation and hydrodynamics, a new chemical equilibrium is calculated in a second step. This iterative
process continues until convergence. We did not not investigate which changes in the spectra were
introduced specifically by using the new ACES equation of state.

\subsubsection{Element abundances}
\label{section:abundances}
\begin{table}
  \caption{Element abundances used in grid, taken from \citet{2009ARA&A..47..481A}. All the abundances
           in the table marked with an m are from meteorites, while the others are
           photospheric.}
  \label{table:abundances}
  \centering                                      %
  \begin{tabular}{llcc|llcc}
    \hline \hline
    Z  & Elem. & Abund. &   & Z  & Elem. & Abund  & \\ \hline
    1  & H     & 12.00  &   & 44 & Ru    & 1.75   &   \\
    2  & He    & 10.93  &   & 45 & Rh    & 1.06   & m  \\
    3  & Li    & 3.26   & m & 46 & Pd    & 1.65   & m  \\
    4  & Be    & 1.38   &   & 47 & Ag    & 1.20   & m  \\
    5  & B     & 2.79   & m & 48 & Cd    & 1.71   & m  \\
    6  & C     & 8.43   &   & 49 & In    & 0.76   & m  \\
    7  & N     & 7.83   &   & 50 & Sn    & 2.04   &   \\
    8  & O     & 8.69   &   & 51 & Sb    & 1.01   & m  \\
    9  & F     & 4.56   &   & 52 & Te    & 2.18   & m  \\
    10 & Ne    & 7.93   &   & 53 & I     & 1.55   & m  \\
    11 & Na    & 6.24   &   & 54 & Xe    & 2.24   &   \\
    12 & Mg    & 7.60   &   & 55 & Cs    & 1.08   & m  \\
    13 & Al    & 6.45   &   & 56 & Ba    & 2.18   &   \\
    14 & Si    & 7.51   &   & 57 & La    & 1.10   &   \\
    15 & P     & 5.41   &   & 58 & Ce    & 1.58   &   \\
    16 & S     & 7.12   &   & 59 & Pr    & 0.72   &   \\
    17 & Cl    & 5.50   &   & 60 & Nd    & 1.42   &   \\
    18 & Ar    & 6.40   &   & 62 & Sm    & 0.96   &   \\
    19 & K     & 5.08   & m & 63 & Eu    & 0.52   &   \\
    20 & Ca    & 6.34   &   & 64 & Gd    & 1.07   &   \\
    21 & Sc    & 3.15   &   & 65 & Tb    & 0.30   &   \\
    22 & Ti    & 4.95   &   & 66 & Dy    & 1.10   &   \\
    23 & V     & 3.93   &   & 67 & Ho    & 0.48   &   \\
    24 & Cr    & 5.64   &   & 68 & Er    & 0.92   &   \\
    25 & Mn    & 5.43   &   & 69 & Tm    & 0.10   &   \\
    26 & Fe    & 7.50   &   & 70 & Yb    & 0.92   & m  \\
    27 & Co    & 4.99   &   & 71 & Lu    & 0.10   &   \\
    28 & Ni    & 6.22   &   & 72 & Hf    & 0.85   &   \\
    29 & Cu    & 4.19   &   & 73 & Ta    & -0.12  & m  \\
    30 & Zn    & 4.56   &   & 74 & W     & 0.65   & m  \\
    31 & Ga    & 3.04   &   & 75 & Re    & 0.26   & m  \\
    32 & Ge    & 3.65   &   & 76 & Os    & 1.40   &   \\
    33 & As    & 2.30   & m & 77 & Ir    & 1.38   &   \\
    34 & Se    & 3.34   & m & 78 & Pt    & 1.62   & m  \\
    35 & Br    & 2.54   & m & 79 & Au    & 0.80   & m  \\
    36 & Kr    & 3.25   &   & 80 & Hg    & 1.17   & m  \\
    37 & Rb    & 2.36   & m & 81 & Tl    & 0.77   & m  \\
    38 & Sr    & 2.87   &   & 82 & Pb    & 2.04   & m  \\
    39 & Y     & 2.21   &   & 83 & Bi    & 0.65   & m  \\
    40 & Zr    & 2.58   &   & 90 & Th    & 0.06   & m  \\
    41 & Nb    & 1.46   &   & 92 & U     & -0.54  & m  \\
    42 & Mo    & 1.88   &   &    &       &        &   \\
    \hline
  \end{tabular}
\end{table}
The element abundances are scaled solar abundances taken from \citet{2009ARA&A..47..481A}, where  
the abundances in both the photosphere and in meteorites are given. With the 
exception of Li, which is destroyed in the solar interior, and volatile elements that are depleted 
in meteorites, both sets of abundances agree within the errors; apart from 
the cases mentioned above, we have therefore chosen the measurements with respective smaller 
uncertainties (see Table~\ref{table:abundances}) to best reproduce the proto-solar elemental 
composition \citep{2003ApJ...591.1220L}.

\subsection{Vacuum wavelengths}
All our spectra are provided at vacuum wavelengths $\lambda_{\mathrm{vac}}$. If air wavelengths
$\lambda_{\mathrm{air}}$ are required, one can use the following equation taken from \cite{1996ApOpt..35.1566C}, 
where it is applied for $\lambda > 2\,000\,\mbox{\AA}$ only
\begin{equation}
  \lambda_{\mathrm{air}}  =  \frac{\lambda_{\mathrm{vac}}}{f},
\end{equation}
with
\begin{eqnarray}
  \sigma_2 & = & \left( \frac{10^4}{\lambda_{\mathrm{vac}}} \right)^2, \\
  f        & = & 1.0 +  \frac{0.05792105}{238.0185 - \sigma_2} + \frac{0.00167917}{57.362 - \sigma_2}.
\end{eqnarray}

\subsection{Interpolation}
A few spectra are missing in the grid where \phx\ was unable to compute the structure of
the atmosphere for various reasons. Where possible, we provide interpolated spectra.

The interpolation is done point-wise, i.e. every wavelength point is interpolated independently
using a cubic spline created from neighbouring spectra. The interpolation is done along the \teff\
axis only.

The difference in flux between an interpolated spectrum and one calculated with \phx\ is usually less
than 1\%, but increases for lower temperatures and can reach up to 10\% for the coolest stars in the grid.

\section{Quality of the spectra}
Many changes have been implemented with respect to previous \phx\ models. For example,
the more detailed treatment of the chemical equilibrium in the new EOS
strongly affects the stellar structure and results in different line and molecular band strengths, 
which can introduce significant differences in comparison to older \phx\ model spectra,
especially for M stars as discussed below.
But major changes can also be expected from the new list of element abundances and the new 
parametrisations for the mixing length and the micro-turbulence. Consequentially we observed some 
significant differences between spectra from previous \phx\ grids and from this one.

We compared some temperature profiles with those of the original models that have been
used as a starting point for the new library. For $\tau \lessapprox 1$ they match very well and we
only see differences for $\tau \gg 1$, which are irrelevant for the morphology of the final spectrum.

\subsection{M stars}
\begin{figure}
  \resizebox{\hsize}{!}{\includegraphics[width=\textwidth]{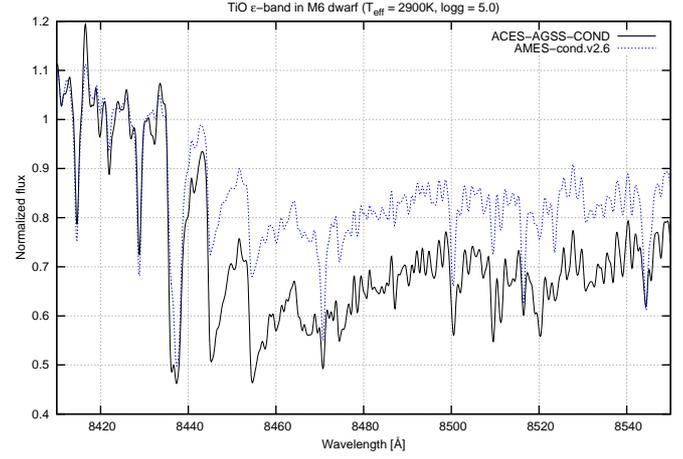}}
  \caption{New \phx\ ACES model spectra (full line) compared to
\phx\ AMES v2.6 spectra (dotted line); both models are calculated for a
typical M6 main sequence star ($\teff = 2\,900$\,K, $\log{g} =
5.0$) and smoothed to an effective resolution of $R=10\,000$. The
ACES model spectra produce significantly deeper TiO bands than earlier
\phx\ models.}
  \label{figure:TiO}
\end{figure}

In Fig.~\ref{figure:TiO} we show a comparison between
model spectra from our new \phx\ grid and from a previous one using
the old EOS AMES version 2.6 at the region around the TiO $\epsilon$-band. In both cases, the
Cond models are used. We show spectra at effective temperatures of
$\teff = 2\,900\,\mathrm{K}$ and $\log{g} = 5.0$, typical values for an M6
main sequence star. The TiO bands are known as robust temperature
indicators, because they only weakly depend on gravity but are very
sensitive to temperature \citep{2004ApJ...609..854M,
2005AN....326..930R}. In the new grid we now see TiO $\epsilon$-bands that
are significantly deeper than those in older \phx\ versions for
identical atmospheric parameters. In this example, the difference in
temperature derived from the two models would be about 200\,K.

\subsection{Comparison with Kurucz models}
\begin{figure}
  \resizebox{\hsize}{!}{\includegraphics{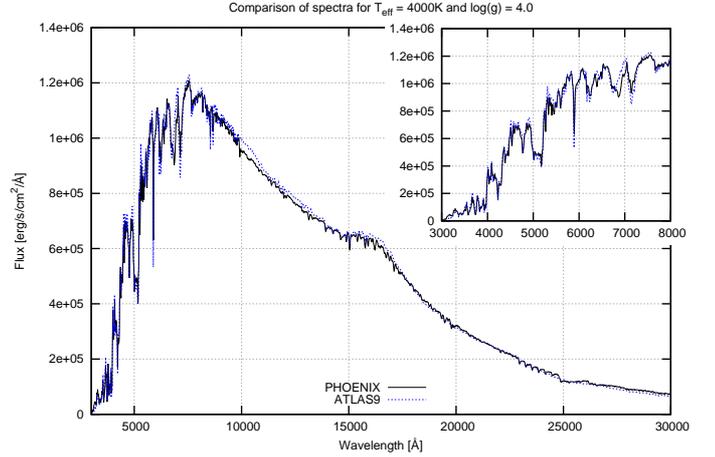}}
  \caption{Comparison between \phx\ (full line) and ATLAS9 (dotted line) model spectra for 
           $\teff=4000\,\mathrm{K}$, $\logg=4.0$ and solar abundances. The \phx\ spectrum
           has been convolved with a Gaussian kernel with a FWHM of 20\,\AA\ for wavelengths
           smaller than 1\,$\mu$m and with a Gaussian with a FWHM of 50\,\AA\ for wavelengths
           larger than 1\,$\mu$m.}
  \label{fig:comp4000K}
\end{figure}
\begin{figure}
  \resizebox{\hsize}{!}{\includegraphics{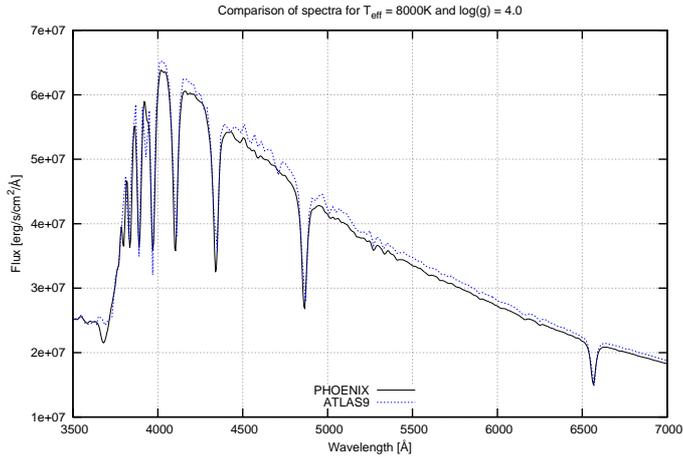}}
  \caption{Comparison between \phx\ (full line) and ATLAS9 (dotted line) model spectra for 
           $\teff=8000\,\mathrm{K}$, $\logg=4.0$ and solar abundances. The \phx\ spectrum
           has been convolved with a Gaussian kernel with a FWHM of 20\,\AA.}
  \label{fig:comp8000K}
\end{figure}
In Figs.~\ref{fig:comp4000K}~and~\ref{fig:comp8000K} the \phx\ synthetic spectra are compared to corresponding ATLAS9 
spectra \citep{2003IAUS..210P.A20C} for two different effective temperatures. Note that they used
the \cite{1998SSRv...85..161G} solar abundances, while we used the more recent abundances by
\cite{2009ARA&A..47..481A}. This will introduce 
systematic differences between the spectra. The \phx\ spectra have been convolved with a Gaussian kernel
with a FWHM of 20\,\AA\ for wavelengths smaller than 1\,$\mu$m and with a Gaussian with a FWHM of 
50\,\AA\ for wavelengths larger than 1\,$\mu$m in order to match the spectral resolution of 
the ATLAS9 spectra better. In general the compared models agree well. For the 8000\,K spectra we observe a 
$\sim$3\% lower flux at every single wavelength point in the \phx\ spectrum than in the ATLAS9 spectrum, 
which is therefore just a difference in total luminosity that can easily be explained by the varying mass in 
our new library.

\subsection{Sun}
\begin{figure}
  \resizebox{\hsize}{!}{\includegraphics{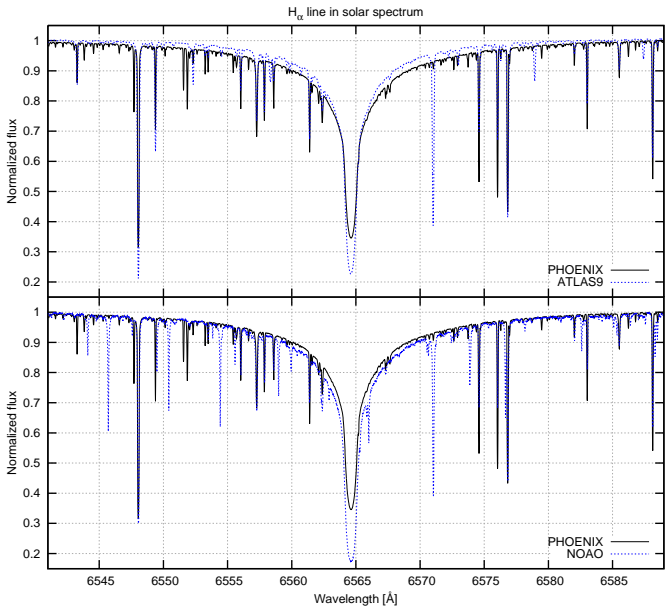}}
  \caption{Comparison between \phx\ (full line) and ATLAS9 (dotted line) model spectra for 
           the Sun in full resolution ($R=500\,000$) in the upper plot. Below the \phx\
           spectrum (full line) is compared to a solar spectrum observed by \cite{2011ApJS..195....6W}.}
  \label{fig:sunHalpha}
\end{figure}
\begin{table}
  \caption{Input parameters for \phx\ and ATLAS9 models for Sun and Vega. Please note that \feh\ in
           the \phx\ models is given by abundances from \cite{2009ARA&A..47..481A}, while ATLAS9
           uses \cite{1989GeCoA..53..197A}.}
  \label{table:paramsVegaSun}
  \centering                                      %
  \begin{tabular}{lllll}
    \hline \hline
    Parameter & \multicolumn{2}{c}{Sun}    & \multicolumn{2}{c}{Vega}    \\ 
              & \phx      & ATLAS9         & \phx      & ATLAS9  \\ \hline
    \teff     & 5778\,K   & 5777\,K        & 9550\,K   & 9550\,K  \\
    \logg     & 4.44      & 4.44           & 3.95      & 3.95 \\
    \feh      & 0.0       & 0.0            & -0.3      & -0.3 \\ \hline
  \end{tabular}
\end{table}
In the upper plot of Fig.~\ref{fig:sunHalpha} a comparison between a \phx\ and an 
ATLAS9\footnote{\url{http://kurucz.harvard.edu/sun.html}} solar spectrum in full resolution is
shown for the region around $H_{\alpha}$. Table~\ref{table:paramsVegaSun} lists the input 
parameters for both. The overall agreement is good, but the line profile is
slightly different: in the \phx\ spectrum the line is not as deep as in the ATLAS9 spectrum, which
is compensated by broader wings. The lower plot of Fig.~\ref{fig:sunHalpha} compares the \phx\ spectrum
to a solar spectrum observed by \cite{2011ApJS..195....6W}. Obviously the real line is even broader and thus
the \phx\ spectrum is a better match than the ATLAS9 one.

Unfortunately, these comparisons also show one of the major drawbacks of synthetic spectra, i.e. they
are missing some of the absorption lines present in the observed spectrum, e.g. the line at 6571\,\AA, which
is present in the ATLAS9 spectrum but not in the \phx\ spectrum.

\begin{table}
  \caption{Comparison of colour indices for different solar spectra.}
  \label{table:coloursSun}
  \centering                                      %
  \begin{tabular}{lccccc}
    \hline \hline
    Spectrum & U-B    & B-V    & V-R    & R-I    & V-I    \\ 
             & [mag]  & [mag]  & [mag]  & [mag]  & [mag]  \\ \hline
    ATLAS9   & 0.069  & 0.654  & 0.443  & 0.324  & 0.767  \\
    \phx     & 0.065  & 0.633  & 0.441  & 0.327  & 0.768  \\
    CALSPEC  & 0.083  & 0.645  & 0.446  & 0.325  & 0.771  \\ \hline
  \end{tabular}
\end{table}
Some colour indices calculated from the ATLAS9 and the \phx\ spectrum are listed in Table~\ref{table:coloursSun}.
In addition, the colours for the solar spectrum from CALSPEC, which is a set of composite stellar 
spectra used for HST calibrations \citep[see e.g.][]{1990AJ.....99.1243T,1997AJ....113.1138C} 
are given. The transmission curves for the used filters have been adapted from 
\cite{1965ApJ...141..923J}\footnote{\url{http://obswww.unige.ch/gcpd/filters/fil08.html}}.

Altogether our solar spectrum seems to be quite accurate, which is gratifying for an all-purpose spectral
library without special optimisations to match the solar spectrum well.

\subsection{Vega}
\begin{figure}
  \resizebox{\hsize}{!}{\includegraphics{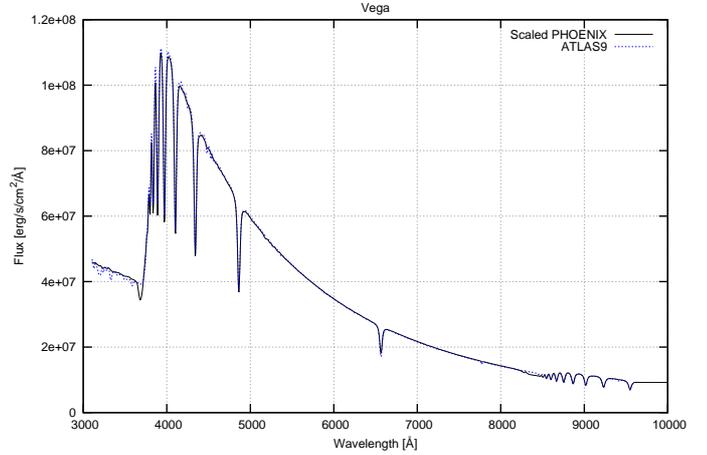}}
  \caption{Comparison between \phx\ (full line) and ATLAS9 (dotted line) model spectra for 
           Vega, both convolved to 20\,\AA\ FWHM. }
  \label{fig:vega}
\end{figure}
In Fig.~\ref{fig:vega} we show a comparison between spectra for Vega, calculated both with \phx\ and 
ATLAS9\footnote{\url{http://kurucz.harvard.edu/stars/vega/}} and the input parameters given
in Table~\ref{table:paramsVegaSun}.
The \phx\ model has been interpolated from the existing grid instead of calculating a new one.
Since the mass of that model differs from the real Vega because of our parametrisation of the 
stellar mass, the flux was scaled to match that of the ATLAS9 model. Further discrepancies are introduced
by the differences in helium abundance, micro-turbulence, etc. Nevertheless, the overall shapes of the spectra
match well.

\begin{table}
  \caption{Comparison of colour indices for different spectra of Vega.}
  \label{table:coloursVega}
  \centering                                      %
  \begin{tabular}{lccccc}
    \hline \hline
    Spectrum & U-B    & B-V    & V-R    & R-I    & V-I    \\ 
             & [mag]  & [mag]  & [mag]  & [mag]  & [mag]  \\ \hline
    ATLAS9   & 0.000  & 0.000  & 0.000  & 0.000  & 0.000   \\
    \phx     & -0.019 & 0.005  & 0.007  & -0.002 & 0.005   \\
    CALSPEC  & 0.027  & 0.006  & 0.006  & 0.008  & 0.014    \\ \hline
  \end{tabular}
\end{table}
Table~\ref{table:coloursVega} lists some colour indices. Since the magnitudes have been calculated
in the VEGAmag system with the ATLAS9 model as reference, all colours are zero for these magnitudes. The results
from our interpolated \phx\ spectrum are in good agreement.

\section{FITS files}

\subsection{Download}
All the spectra presented in this paper are available for
download from our homepage\footnote{\url{http://phoenix.astro.physik.uni-goettingen.de/}}.
Single high-resolution spectra can be downloaded using a web form by specifying temperature, surface
gravity, metallicity and alpha element abundance. In order to keep the amount of storage space
to a minimum, these files contain only the flux; the
wavelength grid must be obtained separately once for all spectra.

Furthermore we provide archives with full sub-grids of the library convolved to more
convenient resolutions:
\begin{enumerate}
 \item $\Delta\lambda=1\mbox{\AA\ }$ in the optical wavelength range from $\lambda=3\,000\,\mbox{\AA\ }$ to $10\,000\,\mbox{\AA}$.
 \item $R=10\,000$, i.e. the resolution of the X-Shooter Spectral Library XSL \citep{2011JPhCS.328a2023C}, with the full 
         wavelength range of X-Shooter, namely $\lambda=3\,000\,\mbox{\AA\ }$ to $24\,800\,\mbox{\AA}$. 
\end{enumerate}
These spectra are over-sampled by a factor of ten, so the 1\,\AA\ grid has a sampling rate of 0.1\,\AA, for example.

\subsection{Naming scheme}
The spectra are provided in a FITS file with the following naming scheme:
\begin{verbatim}
<grid>/<subgrid>/
 (n)lte<Teff><log(g)><subgrid>.<grid>-HiRes.fits
\end{verbatim} 
where \verb|<grid>| is the name of the grid. As for the spectra described in this paper,
this will always be \verb|PHOENIX-ACES-AGSS-COND-2011|. 

The \verb|<subgrid>| describes all the parameters in addition to $\teff$ and 
$\logg$ and is empty when there are none. An example would be \verb|.Alpha=+0.30| 
for an alpha element enhancement of +0.3 dex. Abundances equal to those of the Sun are 
never included, i.e. there is no \verb|.Alpha=+0.00|. Be aware that $\feh$ is given as 
first parameter in the name of the sub-grid, not $Z$.

The filename itself always starts with either \verb|lte| or \verb|nlte|, describing whether
the atmosphere has been calculated in LTE or NLTE.

The effective temperature \verb|<Teff>| is always given by five digits with leading zeros,
if necessary. The surface gravity \verb|<log(g)>| is denoted by its negative with two digits after
the decimal point. In the filename itself the \verb|Z| describing the metallicity is dropped,
since the metallicity is explicitly given for all files. 

Two examples for the naming scheme would be
\begin{verbatim}
* PHOENIX-ACES-AGSS-COND-2011/Z-1.0.Alpha=+0.30/
    lte08000-2.00-1.0.Alpha=+0.30.
    PHOENIX-ACES-AGSS-COND-2011-HiRes.fits
* PHOENIX-ACES-AGSS-COND-2011/Z-0.0/
    lte06800-4.50-0.0.
    PHOENIX-ACES-AGSS-COND-2011-HiRes.fits
\end{verbatim}

\subsection{File content}
The files always contain one single primary extension, which holds the flux of the spectrum
in units of $[\mathrm{erg}/\mathrm{s}/\mathrm{cm}^2/\mathrm{cm}$] on the stellar surface. The
files for the high-resolution spectra contain a FITS keyword WAVE, which holds a reference
to the file that stores the wavelength grid and is saved in the same format as the spectra, but
with wavelength points instead of a flux array.

For the medium-resolution spectra, the wavelength grid is provided by a set of four FITS keywords.
CRVAL1 defines the wavelength at pixel CRPIX1, which is always 1 for our spectra. The
step size is given by CDELT1 and the number of points by NAXIS1. The type of the wavelength 
grid is defined by the CTYPE1 keyword --- for this grid it will always be one of the two
possible values for vacuum wavelengths, i.e. WAVE or WAVE-LOG, where the latter denotes
a logarithmic scale. With this information the whole wavelength grid can be calculated easily.

\subsection{Additional FITS keywords}
In addition to the default FITS keywords, we added some more for documenting the atmospheric parameters
for the spectrum.

The basic parameters are given as
\begin{itemize}
  \item PHXTEFF: [K] effective temperature,
  \item PHXLOGG: [cm/s$^2$] log (surface gravity),
  \item PHXM\_H: [M/H] metallicity (rel. sol. - Asplund et al 2009),
  \item PHXALPHA: [a/M] alpha element enhancement.
\end{itemize}
Additional parameters for the atmosphere are
\begin{itemize}
  \item PHXDUST: [T/F] Dust in atmosphere, always $F$ for this grid,
  \item PHXXI\_L: [km/s] micro-turbulent velocity for LTE lines,
  \item PHXXI\_M: [km/s] micro-turbulent velocity for molec lines,
  \item PHXXI\_N: [km/s] micro-turbulent velocity for NLTE lines,
  \item PHXMASS: [kg] Stellar mass,
  \item PHXREFF: [cm] Effective stellar radius,
  \item PHXLUM: [W] Stellar luminosity,                
  \item PHXMXLEN: Mixing length.
\end{itemize}
Furthermore, we included some information about the \phx\ run:
\begin{itemize}
  \item PHXBUILD: Build date of Phoenix,
  \item PHXVER: Phoenix version,
  \item DATE: [local] finishing date of model calculation,
  \item PHXEOS: Equation of state.
\end{itemize}
As mentioned before, there are models within the grid for which we manually disabled
the convection. For those we added another keyword PHXCONV with the value F.
For the high-resolution spectra there is also a second FITS extension in the file
containing a binary table listing all the element abundances that have been
used in the model.

The files containing interpolated spectra only include the date of their creation and 
those FITS keywords from the lists above that specify the basic stellar parameters 
($\teff$, $\logg$, $\feh$, and $\afe$).
Furthermore, for these spectra we added the keyword INTERPOL with the value T.

\section{Summary}
The presented \phx\ grid contains a comprehensive set of synthetic spectra
that allows a detailed analysis of observed spectra for a wide range of applications. An extension taking into account NLTE treatment of 
important elements up to $25\,000\,\mathrm{K}$ is a work in progress.
Both resolution and wavelength range match existing and upcoming state-of-the-art instruments.

\begin{acknowledgements}
We warmly thank the anonymous referee for his/her comments that helped to improve this publication significantly.
Derek Homeier has been funded under the Labex Lio programme of the Universit\'e Claude Bernard Lyon 1,
the Agence Nationale de la Recherche (ANR), and the Programme National de Physique Stellaire
(PNPS) of CNRS (INSU).
Numerical calculations were performed at the  {\sl Gesellschaft f{\"u}r
Wissenschaftliche Datenverarbeitung G{\"o}ttingen}.

\end{acknowledgements}

\bibliographystyle{aa} 
\bibliography{paper} 

\end{document}